\documentclass[pre,floatfix,longbibliography,superscriptaddress,twocolumn,final,notitlepage]{revtex4-2}

\RequirePackage{amsmath}
\RequirePackage{hyperref} 
\RequirePackage{amssymb}
\RequirePackage{graphicx}
\RequirePackage{bbm}
\RequirePackage{xcolor}
\usepackage{graphicx}
\usepackage{subcaption}
\usepackage{amsmath}
\usepackage{natbib}
\usepackage{caption}
\usepackage[nameinlink,capitalize]{cleveref}

\def\andname{\hspace*{-0.5em}}
\setcitestyle{round}

\newcommand{\be}{\begin{equation}}
\newcommand{\ee}{\end{equation}}
\newcommand{\bea}{\begin{eqnarray}}
\newcommand{\eea}{\end{eqnarray}}
\newcommand{\ra}{\rightarrow}
\newcommand{\f}[2]{\frac{#1}{#2}}
\newcommand{\avg}[1]{\left\langle #1 \right\rangle} 

\newcommand{\ccup}[1]{\left\{#1\right\}}

\newcommand{\bup}[1]{\left(#1\right)}
\newcommand{\rup}[1]{\left[#1\right]}
\newcommand{\pois}{\textrm{Pois}}
\newcommand{\dmax}{D}

\usepackage{algorithm,algorithmic}
\usepackage{amsmath,fixmath}
\usepackage[algoruled,algo2e]{algorithm2e}
\usepackage{fixmath}
\usepackage{setspace}
\usepackage{float}

\usepackage{booktabs}
\usepackage{adjustbox}
 \renewcommand{\thetable}{\arabic{table}}
 
\definecolor{shadecolor}{gray}{0.9}
\usepackage{tabu}

\newcommand{\MC}[1]{\textcolor{magenta}{#1}}

\usepackage[normalem]{ulem} 
\usepackage{soul}

\newcommand{\mt}{\mbox{{\small MULTITENSOR}}}
\newcommand{\gmt}{\mbox{{\small Graph-MT}}}
\newcommand{\hymt}{\mbox{{\small Hypergraph-MT}}}
\newcommand{\pwmt}{\mbox{{\small Pairs-MT}}}
\newcommand{\chodrow}{\mbox{{\small Hypergraph AON-MLL}}}
\newcommand{\SC}{\mbox{{\small Spectral Clustering}}}

\usepackage{enumitem,amssymb}
\newlist{todolist}{itemize}{2}
\setlist[todolist]{label=$\square$}
\newlist{todolist_done}{itemize}{2}
\setlist[todolist_done]{label=$\blacksquare$}
\usepackage{pifont}
\begin{document}

\title[Inference of hyperedges and overlapping communities in hypergraphs]{Inference of hyperedges and overlapping communities in hypergraphs}

\author{Martina Contisciani}
	\affiliation{Max Planck Institute for Intelligent Systems, Cyber Valley, 72076 Tübingen, Germany}

\author{Federico Battiston}
	\affiliation{Department of Network and Data Science, Central European University, 1100 Vienna, Austria}

\author{Caterina De Bacco}
	\email{caterina.debacco@tuebingen.mpg.de}
	\affiliation{Max Planck Institute for Intelligent Systems, Cyber Valley, 72076 Tübingen, Germany}

\begin{abstract} 
Hypergraphs, encoding structured interactions among any number of system units, have recently proven a successful tool to describe many real-world biological and social networks. Here we propose a framework based on statistical inference to characterize the structural organization of hypergraphs. The method allows to infer missing hyperedges of any size in a principled way, and to jointly detect overlapping communities in presence of higher-order interactions. Furthermore, our model has an efficient numerical implementation, and it runs faster than dyadic algorithms on pairwise records projected from higher-order data. We apply our method to a variety of real-world systems, showing strong performance in hyperedge prediction tasks, detecting communities well aligned with the information carried by interactions, and robustness against addition of noisy hyperedges. Our approach illustrates the fundamental advantages of a hypergraph probabilistic model when modeling relational systems with higher-order interactions.
\end{abstract}

\maketitle

\section*{Introduction} 
Over the past twenty years, networks have allowed to map and characterize the architecture of a wide variety of relational data, from social and technological systems to the human brain~\cite{boccaletti2006complex}. Despite their success, traditional graph representation are unable to provide a faithful representation of the patterns of interactions occurring in the real-world~\cite{lambiotte2019networks}. Collections of nodes and links--networks--can only properly encode dyadic relations. Yet, in the last few years systems as diverse as cellular networks~\cite{klamt2009hypergraphs}, structural and functional brain networks~\cite{petri2014homological,giusti2016two}, social systems~\cite{benson2016higher}, ecosystems~\cite{grilli2017higher}, social image search engines~\cite{gao2012visual}, human face-to-face interactions~\cite{cencetti2021temporal} and collaboration networks~\cite{patania2017shape}, have shown that a large fraction of interactions occurs among three or more nodes at a time. 
These higher-order systems are hence best described by different mathematical frameworks such as hypergraphs~\cite{berge1973graphs}, where hyperedges of arbitrary dimensions may encode structured relations among any number of system units~\cite{battiston2020networks, torres2021why, battiston2022higher}.
Interestingly, providing a higher-order description of the system interactions has been shown to lead to the emergence of new collective phenomena~\cite{battiston2021physics} in diffusive~\cite{schaub2020random,carletti2020random}, synchronization~\cite{bick2016chaos,skardal2020higher,millan2020explosive, lucas2020multiorder,gambuzza2021stability}, spreading~\cite{iacopini2019simplicial,chowdhary2021simplicial,neuhauser2020multibody}  and evolutionary~\cite{alvarez2021evolutionary} processes.

To properly describe the higher-order organization of real-world networks, a variety of growing~\cite{kovalenko2021growing,millan2021local} and equilibrium models, such as generalized configuration models~\cite{courtney2016generalized,young2017construction,chodrow2020configuration} have been proposed. Tools from topological data analysis have allowed to obtain insights into the higher-order organization of real-world networks~\cite{patania2017topological,sizemore2019importance}, and methods to infer higher-order interactions from pairwise records have been suggested~\cite{young2021hypergraph}. Finally, several powerful network metrics and ideas have been extended beyond the pair, from higher-order clustering~\cite{benson2018simplicial}, spectral methods~\cite{krishnagopal2021spectral} and centrality~\cite{benson2019three, tudisco2021node} to motifs~\cite{lotito2022higher} and network backboning~\cite{musciotto2021detecting}.

Despite a few recent contributions~\cite{wolf2016advantages,vazquez2009finding,carletti2021random,eriksson2021choosing, chodrow2021generative,chodrow2022nonbacktracking,zhou2006learning}, how to define and identify the mesoscale organization of real-world hypergraphs is still a largely unexplored topic. Here, we propose a new principled method to extract higher-order communities based on statistical inference. More broadly, our approach is that of generative models, which incorporate a priori community structure by means of latent variables, inferred directly from the observed interactions~\cite{ball2011efficient, debacco2017community,goldenberg2010}. Beyond its efficient numerical implementation, our model has several desirable features. It detects overlapping communities, an aspect that is missing in current approaches of community detection in hypergraphs and that is arguably better representative of scenarios where nodes are expected to belong to multiple groups. It also provides a natural measure to perform link prediction tasks, as it outputs the probability that a given hyperedge exists between any subset of nodes. Similarly, it allows to generate synthetic hypergraphs with given community structure, an ingredient that can be given in input or learned from data. Moreover, our explicit higher-order approach is not only more grounded theoretically, but also more efficient than applying graph algorithms to higher-order data projected into pairwise records.\\
We apply our method to a variety of real-world systems, showing that it recovers communities more robustly against noisy addition of large hyperedges than methods on projected pairwise data, it achieves high performance in predicting missing hyperedges, and it allows to determine the influence of hyperedge size in such prediction tasks. We also illustrate how our higher-order approach detects communities that are more aligned with the information carried by hyperedges than what is recorded by node attributes. 
Through these examples, we illustrate how a principled higher-order probabilistic approach can shed light on the role that higher-order interactions
play in real-world complex systems.

\section*{Model}
Here, we introduce \hymt\text{}, a probabilistic generative model for hypergraphs with mixed-membership community structure. Based on a statistical inference framework, our model provides a principled, efficient and scalable approach to extract overlapping communities in networked systems characterized by the presence of interactions beyond the pair.

At its core, our approach assumes that nodes belong to different groups in different amounts, as specified by a set of membership vectors. These memberships then determine the probability that any subset of nodes is connected with a hyperedge. We denote a hypergraph with $N$ nodes $\mathcal{V}=\ccup{i_{1},\dots,i_{N}}$ and $E$ hyperedges $\mathcal{E}=\ccup{e_{1},\dots,e_{E}}$ as $\mathcal{H}(\mathcal{V},\mathcal{E})$. Mathematically, this can be represented as an adjacency tensor $A$ with entries $A_{i_{1},\dots,i_{d}}$ equal to the weight of a $d$-dimensional interaction between the nodes $i_{1},\dots,i_{d}$. For instance, for contact interactions, $A_{i_{1},\dots,i_{d}}$ could be the number of times that nodes $i_{1},\dots,i_{d}$ were in close contact together.

Given these definitions, we can specify the likelihood of observing the hypergraph given a set of latent variables $\theta$, which include the membership vectors. This relies on modelling $P(A_{i_{1},\dots,i_{d}}|\theta)$, the probability of observing a hyperedge given $\theta$. We model this probability as:
\be\label{eqn:eq1}
P(A_{i_{1},\dots,i_{d}} | \theta) = \pois(A_{i_{1},\dots,i_{d}}; \lambda_{i_{1},\dots,i_{d}}) \quad,
\ee
where $ \lambda_{i_{1},\dots,i_{d}} = \sum_{k_{1},\dots,k_{d}} u_{i_{1}k_{1}}\dots u_{i_{d}k_{d}} w_{k_{1},\dots, k_{d}}$. 
The set of latent variables is defined by $\theta= (u,w)$, where $u$ is a $N \times K$-dimensional community membership matrix and $w$ is an affinity tensor, which captures the idea that an interaction is more likely to exist between nodes of compatible communities. If only pairwise interactions exist, the affinity matrix has dimension  $K\times K$. Therefore, the problem reduces to the traditional network case and can be efficiently solved~\cite{debacco2017community}. When higher-order interactions are present, the dimension of the affinity tensor $w$ can become arbitrarily large depending on the size $d_e$ of a hyperedge $e$, i.e., the number of nodes present in it. In fact, $w$ has as many entries as all the possible $d_e$-way interactions between all $K$ groups. For instance, in a hypergraph with only 2-way and 3-way interactions, we have $w=[w^{(2)},w^{(3)}]$ with $w^{(2)}$ of dimension $K\times K$ and $w^{(3)}$ of dimension $K\times K \times K$. 

The question is thus how to reduce the dimension of $w$. A relevant choice that overcomes these problems is that of assortativity~\cite{fortunato2016community},
implying that a hyperedge is more likely to exist when all nodes in it belong to the same group. This captures well situations where homophily, the tendency of nodes with similar features to be connected to each other,  plays a role, as observed in social or biological networks~\cite{asikainen2020cumulative,debacco2017community}. Mathematically, the only non-zero elements of $w$ are the ``diagonal'' ones, that is:
\be\label{eqn:w_ass}
w_{k_1,\dots, k_d} = \delta_{k_{1},\dots, k_{d}}w_{k_{1},\dots, k_{d}}\quad.
\ee 
With this, we obtain a matrix $w$ of dimension $D \times K$, where $D= \max_{e \in \mathcal{E}}d_e$ is the maximum hyperedge size in the dataset. In principle, one could envisage other ways to restrict $w$ to control its dimension. However, we found that the choice in \cref{eqn:w_ass} provides a natural interpretation, results in good prediction performance on both real and synthetic datasets, and is computationally scalable. A similar problem of dimensionality reduction has been tackled in~\cite{chodrow2021generative}, which investigated the more constrained case of hard-membership models.

Putting all together, we model the likelihood of the hypergraph as:
\begin{align}
P(A|\theta) &= \prod_{e \in \Omega} e^{-\lambda_{e}}\,\f{ \lambda_{e}^{A_{e}}}{A_{e}!}\quad,\label{eqn:lik}\\
\text{with} \quad \lambda_e &= \sum_k w_{d_ek} \, \prod_{i \in e} u_{ik}  \quad,
\end{align}
where $\Omega = \ccup{e | e \subseteq \mathcal{V}, d_e\geq 2}$ is the set of all potential hyperedges. In practice, we can reduce this space by considering only the possible hyperedges of a certain size lower or equal than the maximum observed size $D$. In \cref{eqn:lik} we assumed conditional independence between hyperedges given the latent variables, a standard assumption in these types of models.
Such a condition could in principle be relaxed following the approaches of Refs.~\cite{safdari2021generative,contisciani2021community,safdari2022reciprocity}, we do not explore this here.


\begin{table*}[!htp]
    \centering
    \setlength{\tabcolsep}{6.5pt}
    \resizebox{\textwidth}{!}{%
	\renewcommand{\arraystretch}{1.5}
    \begin{tabular}{{l}*{9}{r}{r}{r}{c}{r}}
    \toprule
         & $N$ & $E$ & $E_G$ & $M$ & $M_G$ & $\avg k$ & s($k$) &  $\avg d$ & s($d$) & $D$ & \% $d=2$ & \% $d>2 \in G$ & $K$ \\
        \midrule
        High school & $327$ & $7,818$ & $5,818$ & $172,035$ & $189,928$ & $55.6$ & $27.1$ & $2.3$ & $0.5$ & $5$ & $70.3\%$ & $88.5\%$ & $9$ \\ 
        Primary school & $242$ & $12,704$ & $8,317$ & $106,879$ & $127,886$ & $127.0$ & $55.2$ & $2.4$ & $0.6$ & $5$ & $61.0\%$ & $87.5\%$ & $11$ \\ 
        Workplace & $92$ & $788$ & $755$ & $9,645$ & $9,831$ & $17.7$ & $8.6$ & $2.1$ & $0.3$ & $4$ & $94.2$\% & $88.2\%$ & $5$ \\ 
        Hospital & $75$ & $1,825$ & $1,139$ & $27,835$ & $32,788$ & $59.1$ & $49.0$ & $2.4$ & $0.6$ & $5$ & $60.7\%$ & $95.1\%$ & $4$ \\ 
        Gene-Disease & $4,642$ & $2,738$ & $55,795$ & $4,131$ & $114,444$ & $1.7$ & $3.6$ & $5.8$ & $5.2$ & $25$ & $32.4\%$ & $0.6\%$ & $25$ \\ 
        Justice & $38$ & $2,826$ & $264$ & $15,040$ & $190,790$ & $366.7$ & $203.6$ & $4.9$ & $1.7$ & $9$ & $7.6\%$ & $81.8\%$ & $2$ \\ 
        House bills & $1,494$ & $41,362$ & $360,086$ & $47,212$ & $2,451,751$ & $245.8$ & $251.6$ & $8.9$ & $6.6$ & $24$ & $18.5\%$ & $2.1\%$ & $2$ \\ 
        Senate bills & $293$ & $19,872$ & $22,157$ & $27,300$ & $732,561$ & $482.0$ & $396.9$ & $7.1$ & $5.4$ & $24$ & $16.5\%$ & $14.8\%$ & $2$ \\ 
        House committees & $289$ & $106$ & $2,535$ & $111$ & $4,312$ & $0.7$ & $2.0$ & $8.6$ & $3.6$ & $18$ & $0.9\%$ & $0.0\%$ & $2$ \\
        Senate committees & $282$ & $275$ & $12,761$ & $289$ & $41,008$ & $16.2$ & $12.6$ & $16.6$ & $6.0$ & $25$ & $0.0\%$ & $0.0\%$ & $2$ \\
        Walmart & $1,025$ & $3,553$ & $8,029$ & $5,112$ & $13,769$ & $9.8$ & $16.7$ & $2.8$ & $1.2$ & $11$ & $51.0\%$ & $7.0\%$ & $10$ \\
        Trivago & $6,687$ & $33,963$ & $69,875$ & $40,280$ & $115,533$ & $13.9$ & $13.8$ & $2.7$ & $1.3$ & $26$ & $59.6\%$ & $16.1\%$ & $36$ \\
	\bottomrule          
    \end{tabular}}
	\caption{\label{tab:data} \textbf{Summary of higher-order datasets.} Shown are the number of nodes~($N$), number of hyperedges in the hypergraph~($E$) and in the graph~($E_G$), number of weighted hyperedges in the hypergraph~($M$) and in the graph~($M_G$), mean node degree~($\avg k$), SD of node  degree~(s($k$)), mean hyperedge size~($\avg d$), SD of hyperedge size~(s($d$)), maximum hyperedge size~($D$), percentage of pairwise interactions~(\% $d =2$), percentage of pairwise interactions in the 2-combination set of hyperedges of size bigger than 2 that are already in the graph~(\% $d>2 \in G$), and number of communities~($K$).}
\end{table*}


Having defined \cref{eqn:lik}, the goal is to infer the latent variables $u$ and $w$ given the observed hypergraph $A$. To infer the values of $\theta=(u,w)$, we consider both maximum likelihood estimation (assuming uniform priors on the parameters) and maximum a posteriori estimation (assuming non-uniform priors). The derivations are similar and rely on an efficient expectation-maximization (EM) algorithm~\cite{dempster1977maximum} that exploits the sparsity of the dataset, as detailed in \Cref{app:inference}.

We obtain the following algorithmic updates for the membership vectors:
\be\label{eqn:u}
u_{ik} = \f{\sum_{e \in \mathcal{E}}B_{ie} \,\rho_{ek} }{\sum_{e \in \Omega | i \in e} w_{d_ek} \, \prod_{j \in e | j \neq i} u_{jk}} \quad,
\ee
where $B_{ie}$ is equal to the weight of the hyperedge $e$ to which the node $i$ belongs  (it is an entry of the hypergraph incidence matrix) and $\rho$ is a variational distribution determined in the expectation step of the EM procedure. The numerator of \cref{eqn:u} can be computed efficiently, as we only need the non-zero entries of the incidence matrix, which is typically sparse. Instead, computing the denominator can be prohibitive depending on the value of $D$, the maximum hyperedge size. This is due to the summation over all possible hyperedges in $\Omega$, which requires extracting all  possible combinations $\binom{N}{d}$, for $d=2,\dots,D$. This problem is not present in the case of graphs, as this summation would be over $N^2$ terms at most. This issue clearly highlights the importance of algorithmic efficiency in handling hypergraph data, an aspect that cannot be overlooked to make a model work in practice. 

We propose a solution to this problem that reduces the computational complexity to $O(NDK)$ and makes our algorithm efficient, scalable and applicable in practice. The key is to rewrite the summation over $\Omega$ such that we have an initial value that can be updated at cost $O(1)$ after each update $u_{ik}^{(t)} \ra u_{ik}^{(t+1)}$, which can be done in parallel over $k=1, \dots, K$. This formulation is explained in details in \Cref{app:inference}, where we also show how to edit the updates in \cref{eqn:u} by imposing sparsity (with a proper prior distribution) or by constraining the membership vectors to be probability vectors such that $\sum_k u_{ik}=1$. In both cases, we get a constant term added in the denominators of the updates.

Finally, the updates of the affinity matrix are given by:
\be\label{eqn:w2}
w_{dk} = \f{\sum_{e \in \mathcal{E} | d_{e}=d}A_{e} \,\rho_{ek} }{\sum_{e \in \Omega | d_{e}=d}\, \prod_{j \in e} u_{jk}} \quad .
\ee
These are also computationally efficient to implement and can be updated in parallel. Further details are in \Cref{app:inference}, where we also provide a pseudocode for the whole inference routine. Additionally, \Cref{app:synthetic} presents the validation of our model on synthetic data with ground truth community structure and the comparison against the generative method of \cite{chodrow2021generative} and the spectral method of \cite{zhou2006learning}. \hymt\ shows a strong and increasing performance in recovering communities as the ground truth community structure becomes stronger, similarly to the method of \cite{zhou2006learning}. However, this method is designed to capture hard-membership communities and benefits from having an inference routine similar to the generative process of the synthetic data. In particular, \hymt\ significantly outperforms the competing methods \gmt, \pwmt, and that of~\cite{chodrow2021generative} as soon as the ground truth community structure becomes less noisy. Remarkably, this is observed in synthetic datasets that are generated with a different generating process than that of \hymt. As a consequence, the positive performance of our method confirms the robustness and the reliability of the methodology here introduced.


\begin{figure*}[!htp]
	\centering
	\includegraphics[width=\linewidth]{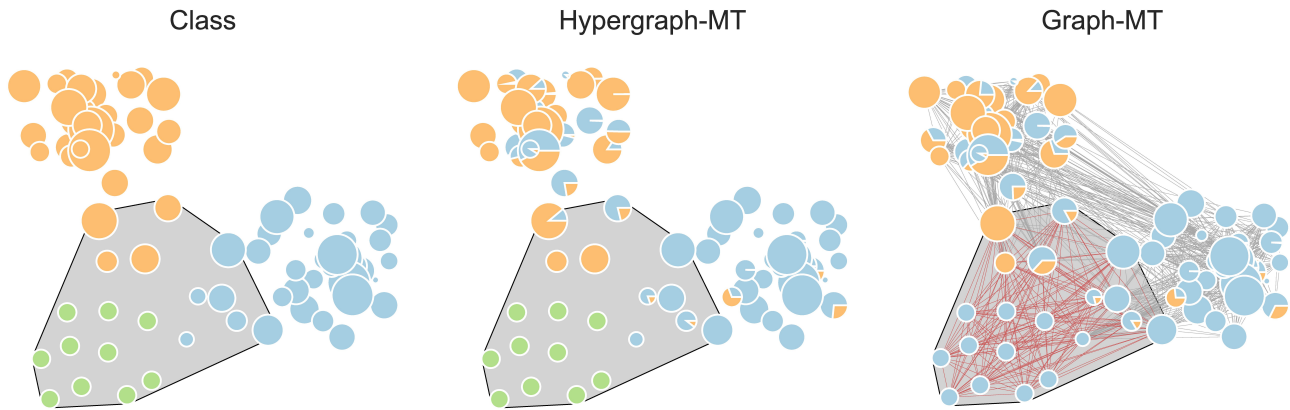}
	\caption{\label{fig:toy_graph} \textbf{The advantage of hypergraph representation: an illustrative example.} The left plot shows a subset of the High school dataset, with nodes belonging to the classes 2BIO1 (light blue) and MP*2 (orange), and ten external guests (green). Node size is proportional to the degree. The gray hyperedge simulates an event, and we omit the other hyperedges for visualization clarity. The central plot displays the partition extracted by \hymt\text{} and on the right we find the partition extracted by \gmt\text{}. In the latter, the gray edges denote the interactions in the graph (obtained by clique expansions) before the event, and the red edges are the interactions added because of the simulated event. This example shows the advantage of using hypergraphs as this representation is more resilient to the addition of a noisy hyperedge and is more robust in detecting communities.}
\end{figure*}


\section*{Results}
We analyze hypergraphs derived from empirical data from various domains. For each one, we report a diverse range of structural properties such as number of nodes, hyperedges and their sizes, as detailed in \Cref{tab:data}. 
Moreover, the datasets provide node metadata, which we use to fix the number of communities $K$, aiming to compare the resulting communities with this additional information. 
For further details on the datasets, see \Cref{app:dataset}.
For each hypergraph, we run \hymt\text{} ten times with different random initialization and select the result with the highest likelihood. For comparison, we run the model on two baselines structures obtained from the same empirical data: a graph obtained from clique expansions of each hyperedge (\gmt), where a hyperedge of size $d$ is decomposed in $\frac{d(d-1)}{2}$ unordered pairwise interactions; a graph obtained using only hyperedges with $d_e=2$ (\pwmt). Notice that running our model on graphs reduces to \mt--the model presented in \cite{debacco2017community}--with an assortative affinity matrix. \MC{} As a remark, we use interchangeably the terms \textit{graph} or \textit{network} to refer to the data with only pairwise interactions, and the term \textit{hypergraph} for the higher-order data.\\

\subsection*{The advantage of using hypergraphs}
The goal of using the two baselines is to assess the advantage (if any) in treating a dataset with higher-order interactions as a hypergraph. Indeed, in practice higher-order data are often reduced to their projected graph, an operation which not only generates a potentially misleading loss of information, but which is also computationally expensive~\cite{wolf2016advantages}. Hence, before evaluating the performance of \hymt \text{} on various datasets, we turn to the following fundamental question: given a dataset of high-order interactions, does a hypergraph representation bring any advantage compared to a simpler graph representation? If the answer is positive, then we should analyze the data with an algorithm that handles hypergraphs. If not, a simpler network algorithm should be enough.

To this end, we analyze four datasets describing human close-proximity contact interactions obtained from wearable sensor data at a high school (High school), a primary school (Primary school), a workplace (Workplace) and a hospital (Hospital). For the analysis, we run the model on the three different structures (hypergraph, clique expansions, and pairwise edges) described above. For each dataset, we compare the inferred partitions with the node metadata that describe either the classes, the departments, or the roles the nodes belong to. We measure closeness to the metadata with the F1-score, a measure for hard-membership classification. It ranges between 0 and 1, where 1 indicates perfect matching between inferred and given partitions. \Cref{tab:socio_f1} shows the performance with the different structures, and both hypergraphs and graphs perform similarly. Notice that the average size of hyperedges in these datasets is around 2.2; thus interactions are mainly pairwise to start with. Moreover, interactions with $d_e >2$ include people who already interact pairwise (see column \% $d > 2 \in G$ in \Cref{tab:data}). Hence, a clique expansion of these is not expected to provide much distinct information from that already present in the pairwise subset of the dataset. Overall, these results suggest that hypergraphs do not bring any additional advantage for these types of datasets, and running a network algorithm would be enough.

\begin{table}[!htp]
	\centering
	\setlength{\tabcolsep}{5pt}
	\resizebox{\linewidth}{!}{%
	\renewcommand{\arraystretch}{1.2}
	\begin{tabular}{{l}*{3}{c}}
	\toprule
				& \hymt	& \gmt		& \pwmt	\\ 
	\midrule
	High school	& 0.757	& 0.776	& 0.755	\\
	Primary school 	& 0.907 	& 0.916	& 0.928	\\
	Workplace	& 0.829	& 0.820	& 0.830	\\
	Hospital		& 0.580	& 0.491	& 0.554	\\
	\bottomrule          
	\end{tabular}}                          
	\caption{\label{tab:socio_f1} \textbf{Comparison of community detection algorithms in human close-proximity contact interactions datasets.} For each dataset, we show the F1-score obtained by comparing a node metadata against the inferred partitions from the hypergraphs (\hymt), the graphs obtained by clique expansions (\gmt), and the graphs given only by the registered pairwise interactions (\pwmt).}
\end{table}

To understand how this assessment may change, we present a toy example built from the High school dataset. 
We select the subset of nodes belonging to two classes (2BIO1 and MP*2 in our example), and we manipulate it by artificially adding a large hyperedge. It simulates an event where ten external people (guests) and a random subset of ten existing nodes are participating. This is represented by the gray hyperedge of dimension 20 in \cref{fig:toy_graph} (left). Here, the green nodes are the external guests, while the blue and orange nodes are the random-selected students from the two classes, respectively. While we only add one hyperedge, its size significantly differs from that of all the other existing hyperedges. In particular, a clique expansion resulting from this additional hyperedge brings in $\binom{20}{2}$  new edges of size $2$ (red in the figure). Hence, we expect this additional information to impact the structure of \gmt\text{} much more than the hypergraph. \Cref{fig:toy_graph} shows that \hymt\text{} is not biased by the presence of this individual large hyperedge, and it well recovers the external guests by assigning zero memberships to them for both classes. Conversely, \gmt\text{} assigns the guests to the blue class. With this toy example, we show a possible scenario where hypergraphs have an advantage, as this representation is more resilient to the addition of a noisy hyperedge and is more robust in detecting communities.

\subsection*{Hyperedge prediction: Analysis of a Gene-Disease dataset}
We now turn our attention to the analysis of a higher-order Gene-Disease dataset, where nodes are genes, and a hyperedge connects genes that are associated with a disease. Here, we focus on the ability of our model to predict missing hyperedges. We measure prediction performance using a cross-validation protocol where hyperedges are divided into train and test sets. The train set is used for parameter estimation, while performance is evaluated on the test set. We compute the area under the receiver-operator curve (AUC), and use the probability assigned by our model of a hyperedge to exist as input scores for this metric. For \gmt\text{}, the probability of a hyperedge to exist is computed as the product of the probabilities that each single edge exists. For details, see \Cref{app:auc}. When evaluating \pwmt, we measure the AUC on the subset of test hyperedges of size 2. To perform a balanced comparison in this case, we also measure the AUC for both \hymt\text{} and \gmt\text{} on this set (pairs), while still training on the whole train set. This provides information on the utility of large hyperedges to predict pairwise interactions. 

\begin{figure}[htbp]
	\centering
	\includegraphics[width=\linewidth]{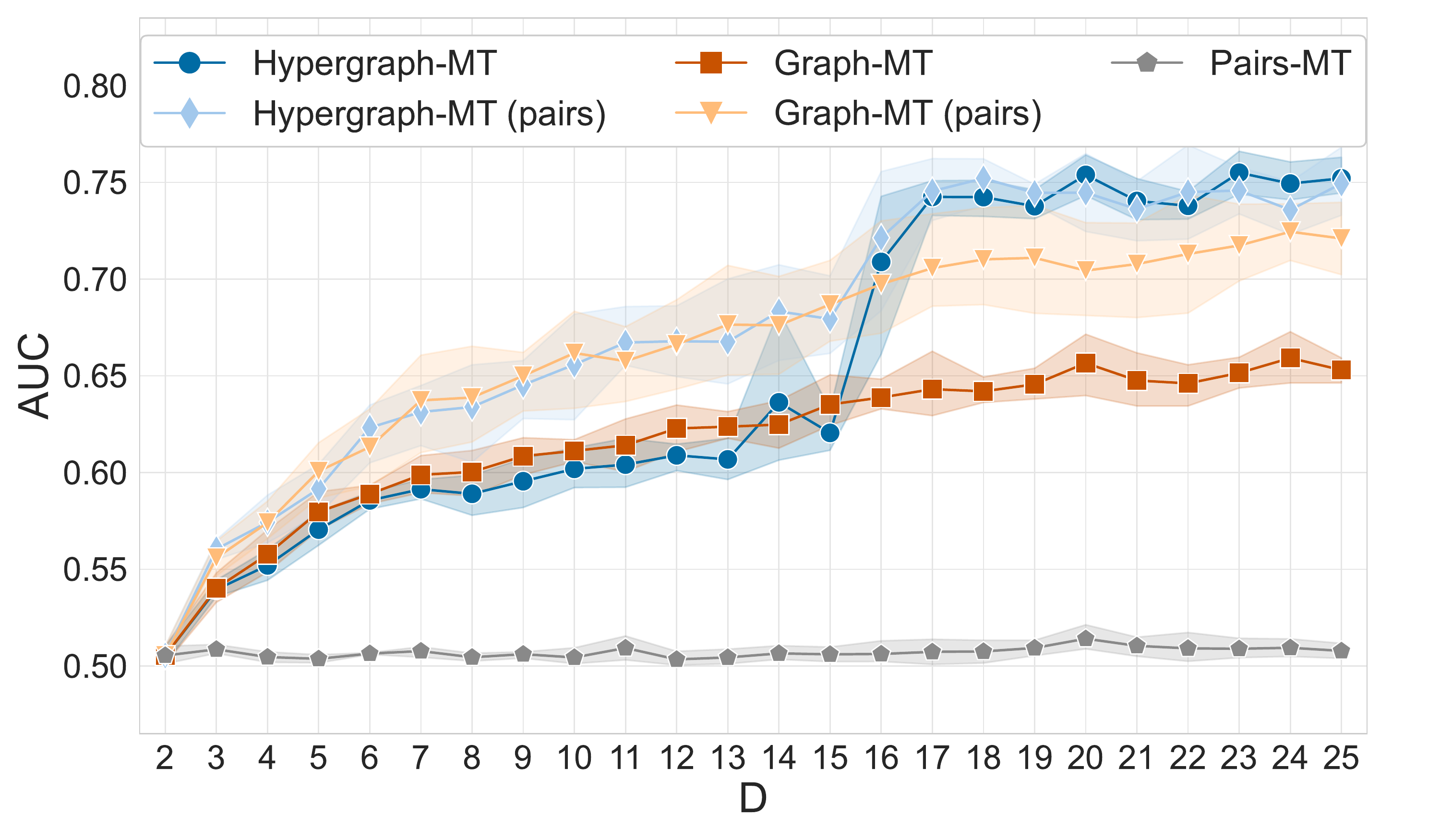}
	\caption{\label{fig:genedisease_auc} \textbf{Critical size for hyperedge prediction in a Gene-Disease dataset.} We measure the AUC by varying the maximum hyperedge size $D$. The results are averages and standard deviations over 5-fold cross-validation test sets, and the baseline for AUC is the random value 0.5. We run the model on the hypergraphs (\hymt), the graphs obtained by clique expansions (\gmt), and the graphs given only by the registered pairwise interactions (\pwmt). To perform a balanced comparison against \pwmt, for \hymt\text{} and \gmt\text{} we additionally measure the AUC on the subset of test hyperedges of size 2 (pairs), while still training on the whole train set. The plot shows the existence of a critical hyperedge size beyond which the higher-order algorithm significantly outperforms alternative methods.}
\end{figure}

We vary the maximum hyperedge size $D$  to show how each method responds to the incorporation of progressively larger edges in terms of prediction tasks. Interestingly, we observe a strong shift in performance around $D=15,16$, where \hymt \text{} significantly outperforms \gmt \text{} and \pwmt\text{} (see \cref{fig:genedisease_auc}). This highlights that hyperedges with larger size carry useful information that cannot be fully captured via clique expansions. This is true regardless of the type of missing edges being predicted (hyperedges or pairs-only). In addition, predictive performance is improved homogeneously across hyperedge sizes in the held-out set. 
Namely, we are not improving just in predicting the pairs-only, as shown by Hypergraph-MT (pairs), but also those of bigger sizes, see \cref{fig:genedisease_auc_prop}. This is where \gmt\text{} fails because the additional information introduced by the clique expansions produces a  much denser graph than the input data that may not be correlated with the true existing hyperedges, thus blurring the observations given in the input. These results not only highlight the ability of our model to predict missing data, but also how the knowledge of large hyperedges helps the prediction of hyperedges of smaller sizes.

\subsection*{Overlapping communities and interpretability: Analysis of a Justice dataset}


\begin{figure*}[!htp]
	\centering
	\includegraphics[width=\linewidth]{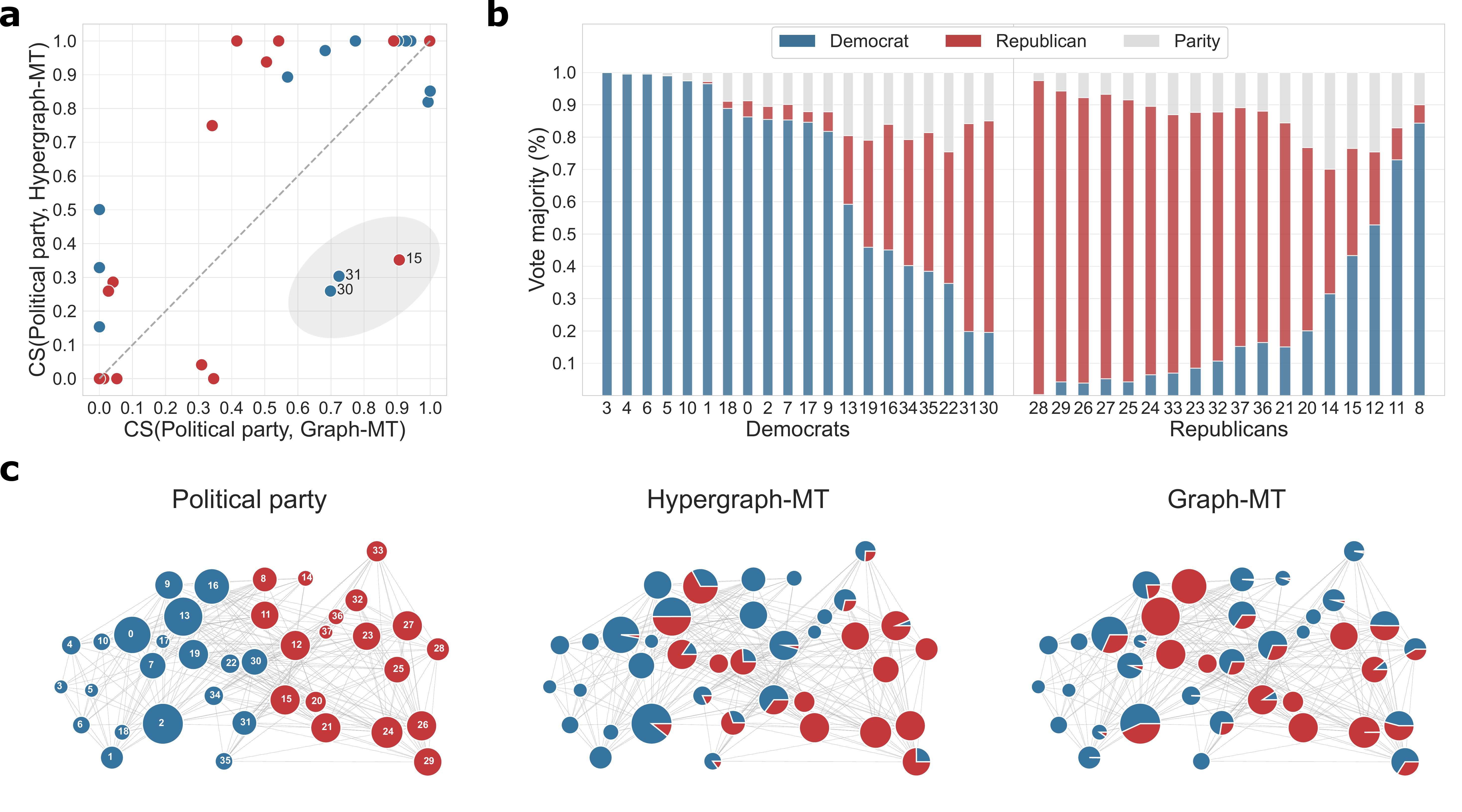}
	\caption{\label{fig:Justice} \textbf{Inference of overlapping communities in a co-voting higher-order dataset of the U.S. Justices.} 
	(\textbf{a}) Point-by-point comparison between the cosine similarities (CS) obtained by \hymt\text{} and \gmt. For each Justice (marker in the plot), we compute the CS between the partitions inferred by the methods and the political party of the Justices, i.e., Democrat (blue) and Republican (red). 
	(\textbf{b}) Vote majority proportion of the hyperedges of each Justice. Every hyperedge is colored based on the majority political party of the Justices involved in it, i.e., either Democratic, Republican, or equally distributed (gray). Then, for every Justice, we extract the percentage of times that they participate in hyperedges of a given majority. 
	(\textbf{c}) Data partition according to the political party (left), and the mixed-membership communities inferred by \hymt\text{} (center) and \gmt (right). Node size is proportional to the degree, node labels are Justice IDs, and the interactions are the edges of the projected graph.}
\end{figure*}


\begin{figure*}[!htp]
	\centering
	\includegraphics[width=\linewidth]{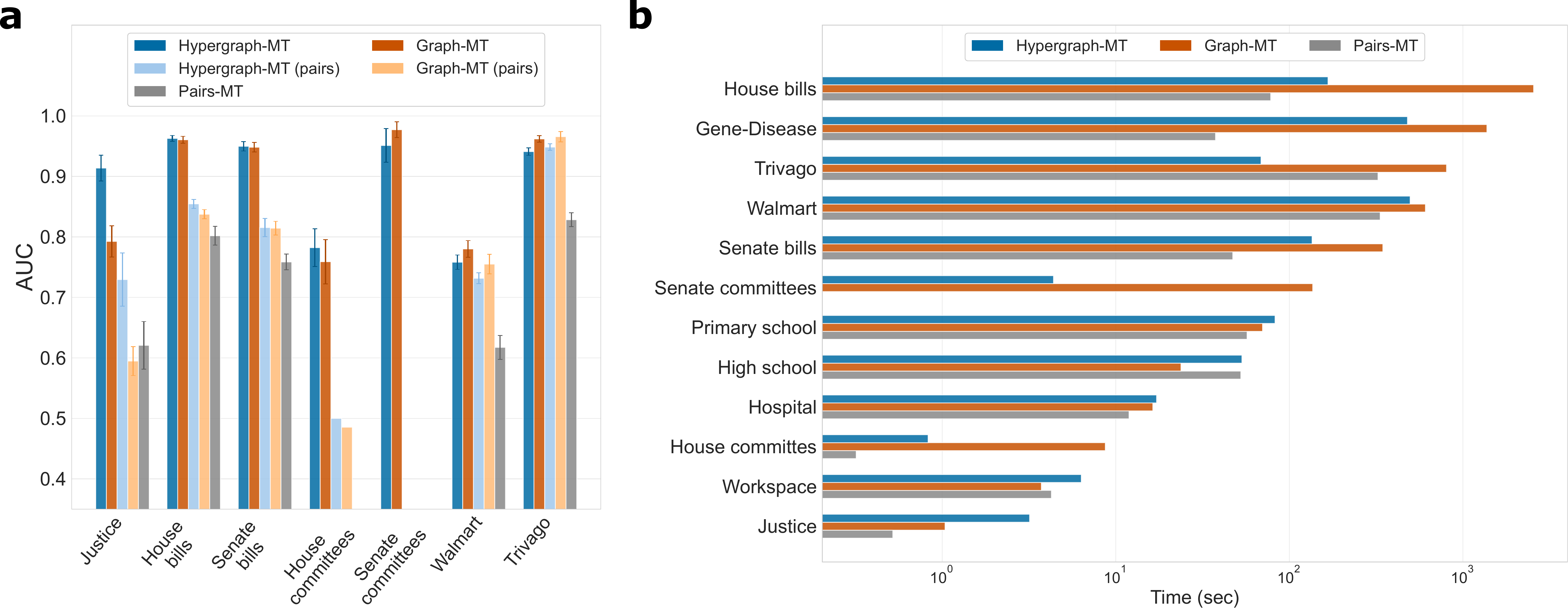}
	\caption{\label{fig:AUCTimeComp} \textbf{Hyperedge prediction performance and computational complexity in higher-order datasets.} (\textbf{a}) The performance of hyperedge prediction is measured with the AUC, whose baseline is the random value 0.5. The results are averages and standard deviations over 5-fold cross-validation test sets. For each dataset, we run the model on the hypergraphs (\hymt), the graphs obtained by clique expansions (\gmt), and the graphs given only by the registered pairwise interactions (\pwmt). To perform a balanced comparison against \pwmt, for \hymt\text{} and \gmt\text{} we additionally measure the AUC on the subset of test hyperedges of degree 2 (pairs), while still training on the whole train set. (\textbf{b}) Computational complexity of \hymt, \gmt, and \pwmt\text{} for the different higher-order datasets. We show the running time for one realization.}
\end{figure*}


Together with hyperedge prediction, \hymt\text{} allows to extract relevant information also on the mesoscale organization of real-world hypergraphs.
As a case study, we analyze a dataset recording all the votes expressed by the Justices of the Supreme Court in the U.S. from 1946 to 2019 case by case. Justices are nodes, and hyperedges connect Justices that expressed the same vote in a given case. The structure of this hypergraph is different from the others analyzed above: it has fewer nodes ($N=38$) but it is denser ($E=2826$), on average a Justice votes $367$ times. Similarly, the graph obtained with clique expansion has substantially fewer edges ($E_G=264$) but with higher weights than the hypergraph. See \cref{tab:data} for details. Examining the communities inferred in these two markedly distinct structures can provide direct insights into the particular aspects captured by a hypergraph formulation.
To this end,  we compare the inferred partitions with the political parties of the Justices, i.e., Democrat or Republican, information provided as node metadata. 
We use the cosine similarity (CS), a metric that measures the distance between vectors, and thus it is better suited to capture mixed-membership communities. The CS varies between 0 and 1, where 1 means that the inferred partition matches perfectly the one shown by political affiliation. 
For each node, we compute the CS between its political party and the partitions inferred by \hymt\text{} and \gmt\text{}. \cref{fig:Justice}a shows the point-by-point comparison between the resulted cosine similarities of the two methods. Here, each marker is a Justice and colors represent their political parties. Points above (below) the diagonal represent Justices for which the communities inferred by \hymt\text{} (\gmt\text{}) align better with the political party. In several cases the two models infer memberships that align similarly with political affiliation: upper-right corner, where both models are aligned well, and lower-left corner, where they are both not aligned well. The interesting behavior is shown in the bottom-right area highlighted in gray, containing three Justices whose political affiliations are more closely associated with the communities inferred by \gmt\text{} than those of \hymt\text{}. To investigate these cases, we inspect the information carried by the hyperedges. Specifically, for each hyperedge we measure the majority political party based on the affiliation of the Justices involved in it. For instance, a hyperedge of size 5 made of 4 democrats and 1 republican has a Democratic majority. We also account for ties, when equal numbers of Justices are in both parties.  Then, for each Justice, we extract the percentage of times that they participate in hyperedges of a given majority. This measure indicates the tendency of Justices to vote more often aligned with democrats or republicans, an information summarized in \cref{fig:Justice}b.  We observe Justices that consistently vote with their own party majority (e.g., Justice 3 votes mainly with other democrats, Justice 28 mainly with other republicans), but also cases in which the political party of the Justice is not aligned with the voting behavior expressed by their hyperedges. For example, node 30 (Justice Ruth Bader Ginsburg) is associated with the Democratic Party, but most of her votes align with those of republican Justices. This behavior is captured by \hymt\text{},  which assigns her a membership  more peaked in the community made of republicans and only partially to the one of democrats, as shown in \cref{fig:Justice}c. Instead, \gmt\text{} assigns her mostly to the community of democrats. This mismatch between hypergraph information and political affiliation explains the lower value of cosine similarity in \cref{fig:Justice}a. Similar conclusions can be drawn for node 31 and 15. More generally, the overlapping memberships inferred by \hymt\text{} match more closely the voting behavior of Justices than those inferred by \gmt\text{}, as shown in the pie markers in \cref{fig:Justice}c. \\

In addition to community structure, \hymt\text{} outperforms \gmt\text{} also in the hyperedge prediction task. \cref{fig:AUCTimeComp}a shows how \hymt\text{} achieves higher AUC than \gmt\text{}, in both predicting pairwise and higher-order interactions. This further corroborates the hypothesis that information is lost when decoupling higher-order interactions via clique expansion.
This example illustrates why it is critical to consider hypergraphs when hyperedges contain information that can be lost by clique expansion. It also shows the advantage of considering overlapping communities when nodes' behaviors are nuanced and no clear affiliation to one group is expected. As Supreme Court cases span a wide range of topics, we may expect Justices to exhibit a diversity of preferences (and thus voting behaviors) that cannot be fully captured by a binary political affiliation. Hence, models that consider overlapping communities can provide a variety of patterns that better represents this diversity. Finally, this example also confirms that metadata should be carefully used as ``ground-truth'' communities, thus encouraging a careful exploration of the relationship between node metadata, information contained in the hyperedges and community
structure~\cite{peel2017ground}. 

\subsection*{The computational efficiency of \hymt}

Beyond accuracy, algorithmic efficiency is necessary for a widespread applicability of statistical inference models to large-scale datasets. Hence, we now assess the performance of our model on a variety of systems from different domains, focusing on the analysis of the computational efficiency of \hymt\text{} as compared to alternative approaches.
The higher-order datasets include co-sponsorship and committee memberships data of the U.S. Congress, co-purchasing behavior of customers on Walmart, and clicking activity of users on Trivago (\Cref{tab:data}).
\hymt \text{} and \gmt \text{} perform similarly in terms of predicting missing hyperedges on most of these datasets, as shown in \Cref{fig:AUCTimeComp}a. This suggests that in such cases, the information learned from the clique expansion is similar to that contained in a hypergraph representation. While one may be tempted to conclude that using a dyadic method should be favored in these cases, we argue that predictive performance may not be the only metric to use to make this decision. Indeed, time complexity also plays a role here, as many of these datasets have large hyperedges. While we have extensively discussed the efficiency of \hymt \text{}, one should also consider the cost of running dyadic methods on clique expansion of large data. In fact, this depends on the number of pairs generated in the expansion, a quantity related to both the amount and size of hyperedges. As a result, the size of a graph obtained by clique expansion can become arbitrarily large. For instance, the House bills data results in almost $4 \times 10^5$ edges, as opposed to the $4 \times 10^4$ hyperedges given by the hypergraph representation. This difference of an order of magnitude has a significant impact in terms of computational complexity. In fact, we observe a difference of an order of magnitude also in the running time of the algorithms, as shown in \Cref{fig:AUCTimeComp}b, where we plot the time to run the three methods on each dataset. While for datasets with small hyperedges (e.g., the close-proximity data discussed above) running time is similar for \hymt \text{} and \gmt, we observe significant differences for datasets with larger maximum size $D$, with \hymt \text{} being much faster to run. \hymt \text{} may therefore be the algorithm of choice for large system sizes. See \Cref{app:synthetic} for further results about the computational complexity of the methods on synthetic data with variable size.

\section*{Discussion}
Here we have introduced \hymt, a mixed-membership probabilistic generative model for hypergraphs, which 
 proposes a first way to extract the overlapping community organization of nodes in networked systems with higher-order interactions. In addition to detecting communities, our model provides a principled tool to predict missing hyperedges, thus serving as a quantitative evaluation framework for assessing goodness of fit. This feature is particularly useful in the absence of metadata when evaluating community detection schemes.
 In practice, our model considers an assortative affinity matrix, which  makes its algorithmic implementation highly scalable. The computational complexity is also significantly reduced by an efficient routine to compute expensive quantities at low cost in each update, a problem not present in the case of graphs. We have applied our model to a wide variety of social and biological hypergraphs, discussing accuracy in the hyperedge and community structure inference tasks. Moreover, we have showed that \hymt\text{} outperforms clique expansion methods with respect to running time, making it a suitable solution also for higher-order datasets with large hyperedges.\\
 
Our method has a substantial advantage in systems where hyperedges contain important information that can be lost by considering non-higher-order methods on projected dyadic graphs. For instance, it allows quantifying how maximum hyperedge size impacts performance and unveils the presence of critical sizes beyond which higher-order algorithms may significantly outperform dyadic methods, as shown in a Gene-Disease dataset. \hymt\text{} also has the benefits of being more resilient to the addition of large noisy hyperedges and of being more robust in detecting communities that are more closely aligned with the information carried by hyperedges, as shown in the analysis of the U.S. Justices.\\

There are natural methodological extensions to further expand the range of applications covered by our model. Here we have considered an assortative affinity matrix, but alternative formulations could be considered to target different types of structures. The challenge would be to increase flexibility while keeping the dimensionality of the problem under control. Moreover, our model takes in input hyperedges of one type, but there could be multiple types of ways to connect a subset of nodes. Expanding our approach to these cases would be analogous to extend single-layer networks to multilayer ones. This may be done by suitably defining different types of affinity matrices for each type of high-order interaction, as in~\cite{debacco2017community}. Similarly, 
our model might be extended to extract temporal higher-order communities in the presence of time-varying interactions with memory~\cite{scholtes2014causality, rosvall2014memory}.
Finally, hypergraphs may carry additional information beyond the one contained in hyperedges. This calls for further developments to rigorously incorporate information such as node attributes into the model formulation~\cite{contisciani2020community,newman2016structure}. While here we have focused on analyzing real-world data, our generative model can also be used to sample synthetic data with hypergraph structure. In particular, our model could prove useful for practitioners interested in utilizing synthetic benchmarks of hypergraphs, allowing a better characterization of higher-order topological properties, including simplicial closure~\cite{benson2018simplicial} and higher-order motifs~\cite{lotito2022higher}. Taken together, \hymt\text{} provides a fast and scalable tool for inferring the structure of large-scale hypergraphs, contributing to a better understanding of the networked organization of real-world higher-order systems.

\bibliographystyle{ScienceAdvances}
\bibliography{bibliography}

\section*{Acknowledgements}
The authors thank the International Max Planck Research School for Intelligent Systems (IMPRS-IS) for supporting M.C; C.D.B. and M.C. were supported by the Cyber Valley Research Fund. F.B. acknowledges support from the Air Force Office of Scientific Research under award number FA8655-22-1-7025. The authors thank Philip S. Chodrow and Nate Veldt for useful discussions.

\section*{Author contributions}
M.C. and C.D.B. developed the algorithm and performed the experiments. M.C., F.B., and C.D.B. all conceived the research, analyzed the results and wrote the manuscript. 

\section*{Competing interests}
The authors declare no competing interests.

\section*{Data availability}
The datasets used in the paper are publicly available from their sources listed in the Supplementary Materials. 

\section*{Code availability}
An open-source algorithmic implementation of the model is publicly available and can be found at \url{https://github.com/mcontisc/Hypergraph-MT}.

\newcommand{\beginsupplement}{%
        \setcounter{table}{0}
        \setcounter{figure}{0}
        \setcounter{equation}{0}
        \renewcommand{\thesection}{\Alph{section}}
	\renewcommand{\thesubsection}{\thesection.\arabic{subsection}}
	\renewcommand{\theequation}{S\arabic{equation}}
	\renewcommand{\thetable}{S\arabic{table}}
	\renewcommand{\thefigure}{S\arabic{figure}}
 }

\clearpage
\beginsupplement
\begin{widetext}
\section*{{Supporting Information (SI)}}

\section{Inference of \hymt}\label{app:inference}
\hymt\text{} models the likelihood of the hypergraph $A=\{A_{e}\}_{e \in \mathcal{E}}$ as:
\be
P(A|\theta) = \prod_{e \in \Omega} e^{-\lambda_{e}}\,\f{\lambda_{e}^{A_{e}}}{A_{e}!}\quad,
\ee
where $ \lambda_e = \sum_k w_{d_ek} \, \prod_{i \in e} u_{ik}$.  The set of latent variables is defined by $\theta= (u,w)$, where $u$ is a $N \times K$-dimensional community membership matrix and $w$ is a $D\times K$-dimensional affinity matrix,  where $D= \max_{e \in \mathcal{E}}d_e$ is the maximum hyperedge size in the dataset. Each entry $w_{dk}$ represents the density of hyperedges of size $d$ in the community $k$. Notice, we only consider the assortative regime, to reduce the dimensionality of the affinity tensor $w$. The product runs over $\Omega = \ccup{e | e \subseteq \mathcal{V}, d_e\geq 2}$, that is, the set of all potential hyperedges. In practice, we can reduce this space by considering only the possible hyperedges of a certain size lower or equal than the maximum observed size $D$. For instance, if the maximum size of interactions in a hypergraph is $\dmax=4$, then we should not expect to see hyperedges of size 5, and we can define $\Omega= \ccup{e | e \subseteq \mathcal{V}, 2\leq d_e\leq D}$. 

With this formulation, \hymt\text{} is a mixed-membership probabilistic generative model for hypergraphs. The main intuition behind it is that a hyperedge is more likely to exist between nodes with the same community membership. In fact, hyperedges in which even a single value $u_{ik}=0$ appears, are assigned a null probability. 
The goal is thus to infer the latent variables $u$ and $w$ given the observed hypergraph $A$. \\

We infer the parameters using a maximum likelihood approach. Specifically, we maximize the log-likelihood
\be
L = -\sum_{e \in \Omega} \sum_{k}w_{d_ek} \prod_{i \in e} u_{ik} + \sum_{e \in \mathcal{E}} A_{e}\log \sum_{k}w_{d_ek} \prod_{i \in e} u_{ik} \label{eqn:loglik}
\ee
with respect to $\theta= (u,w)$, where we neglect the factorial term which is independent of the parameters. Because the summation in the logarithm renders the calculations difficult, we employ a variational approximation using Jensen’s inequality, that gives
\be
\mathcal{L}(\rho, \theta) =  -\sum_{e \in \Omega} \sum_{k}w_{d_ek} \prod_{i \in e} u_{ik} +
 \sum_{e \in \mathcal{E}} A_{e} \sum_{k}\rho_{ek}\log \bup{\f{w_{d_ek} \prod_{i \in e} u_{ik}}{\rho_{ek}} } \quad . \label{eqn:logVI}
\ee
For each $e \in \mathcal{E}$, we consider a variational distribution $\rho_{ek}$ over the communities $k$: this is our estimate of the probability that the hyperedge $e$ exists due to the contribution of the community $k$. The equality holds when
\be
\rho_{ek} = \f{w_{d_ek} \prod_{i \in e} u_{ik}}{\sum_{k}w_{d_ek} \prod_{i \in e} u_{ik}} \quad. \label{eqn:rho}
\ee
Maximize \cref{eqn:loglik}, is then equivalent to maximize \cref{eqn:logVI} with respect to both $\theta$ and $\rho$. We estimate the parameters by using an expectation-maximization (EM) algorithm, where at each step one updates $\rho$ using \cref{eqn:rho} (E-step) and then maximizes $\mathcal{L}(\rho, \theta)$ regarding $\theta=(u,w)$ by setting partial derivatives to zero (M-step). This procedure is repeated until the log-likelihood converges. The fixed point is a local maximum, but it is not guaranteed to be the global maximum. Therefore, we perform ten runs of the algorithm with different random initialization for $\theta$, taking the fixed point with the largest value of the log-likelihood.

\subsection{Expectation-Maximization updates}

The derivative in $u_{ik}$ is given by:
\be
\f{\partial  \mathcal{L}}{\partial u_{ik} }= -\sum_{e \in \Omega | i \in e} w_{d_ek} \prod_{j \in e | j \neq i} u_{jk} + \sum_{e \in \mathcal{E} | i \in e}A_{e} \f{\rho_{ek} }{u_{ik}} \quad.
\ee
Setting this to zero, we obtain the updates:
\be\label{eqn:u_app}
u_{ik} = \f{\sum_{e \in \mathcal{E} | i \in e}A_{e} \,\rho_{ek} }{\sum_{e \in \Omega | i \in e} w_{d_ek} \prod_{j \in e | j \neq i} u_{jk}} =  \f{\sum_{e \in \mathcal{E} }B_{ie} \,\rho_{ek} }{\sum_{e \in \Omega | i \in e} w_{d_ek} \prod_{j \in e | j \neq i} u_{jk}} \quad ,
\ee
where $B_{ie}$ is equal to the weight of the hyperedge $e$ to which the node $i$ belongs  (it is an entry of the hypergraph incidence matrix). The numerator of \cref{eqn:u_app} can be computed efficiently, as we only need the non-zero entries of the incidence matrix, which is typically sparse. Instead, computing the denominator can be prohibitive depending on the value of $D$, the maximum hyperedge size. This is due to the summation over all possible hyperedges in $\Omega$, which requires extracting all  possible combinations $\binom{N}{d}$, for $d=2,\dots,D$. 
We propose a solution to this problem that reduces the computational complexity to $O(NDK)$. The key is to rewrite the summation over $\Omega$ such that we have an initial value that can be updated at cost $O(1)$ after one update of $u_{ik}^{(t)} \ra u_{ik}^{(t+1)}$. 
Defining the set of hyperedges of fixed size $d$ as $\Omega^{d}=\ccup{e \in \Omega, d_e=d}$ and  $\bar{\Omega}^{d}_{ik}= \ccup{e \in \Omega^{d}| i \notin e}$, we can write more compactly:
\be\label{eqn:u2}
u_{ik} =  \f{\sum_{e\in \mathcal{E}}B_{ie} \,\rho_{ek} }{\sum_{d=2}^{D} w_{dk}\sum_{e \in \bar{\Omega}_{ik}^{d-1}}  \, \prod_{j \in e } u_{jk}} \quad.
\ee
The idea now is to observe that products like $\sum_{e \in \bar{\Omega}_{ik}^{d-1}}  \, \prod_{j \in e } u_{jk}$ can be written as a function of $\sum_{e \in \Omega^{d-1}}  \prod_{j \in e} u_{jk}$ and $u_{ik}$. The first term depends exclusively on $d$ and $k$, not on a particular $i$. Hence, by isolating these terms from the ones that depend on $u_{ik}$, we only need to update the second terms, without re-computing the first. For instance, for $d=3$, we can write $\sum_{e \in \bar{\Omega}_{ik}^{2}}  \, \prod_{j \in e } u_{jk}=\sum_{j\neq m|j,m\neq i } u_{jk}u_{mk} = \sum_{e \in \Omega^{2}}\prod_{j \in e}u_{jk} - u_{ik}\sum_{j \neq i}u_{jk}$. 

To formalize this, we define a function $\psi(S,k)$ that depends on a set of hyperedges $S$ and community index $k$ as
\be
\psi(S,k)=\sum_{e \in S}\prod_{j \in e}u_{jk}\quad.
\ee
With this definition, we have
\be\label{eqn:psiomega}
\sum_{e \in \Omega^{d}}\prod_{j \in e}u_{jk} = \psi(\Omega^{d},k)=u_{ik}\, \psi(\bar{\Omega}_{ik}^{d-1},k) +  \psi(\bar{\Omega}_{ik}^{d},k) \,,
\ee
valid for $d=1,\dots,D$ and we fix the term $ \psi(\bar{\Omega}_{ik}^{0},k)=1$. Notice that $ \psi(\Omega^{d},k)$ depends solely on $d$ and $k$, and it can be used to compute $\psi(\bar{\Omega}_{ik}^{d},k)$, needed in the denominator of \cref{eqn:u2}. 
The advantage of using this formulation is given by the efficient update procedure. Indeed, when we update an entry $u_{ik}^{(t)} \ra u_{ik}^{(t+1)}$ we can efficiently update $\psi^{(t)}(\Omega^{d},k)$ by simply:
\be
\psi^{(t)}(\Omega^{d},k) = \bup{u_{ik}^{(t)}-u_{ik}^{(t-1)}} \psi^{(t-1)}(\bar{\Omega}^{d-1}_{ik},k) +  \psi^{(t-1)}(\Omega^{d},k) \, ,
\ee
and we can do this in parallel over $k=1,\dots,K$.
These new values can then be used in the denominator of any other update $u_{jk}^{(t-1)} \ra u_{jk}^{(t)}$ as
\be
\psi^{(t)}(\bar{\Omega}^{d}_{jk},k)  = \psi^{(t)}(\Omega^{d},k)  - u^{(t-1)}_{jk}\, \psi^{(t)}(\bar{\Omega}^{d-1}_{jk},k)\quad.
\ee

In practice, one only needs to initialize the values of $ \psi^{(0)}(\Omega^{d},k) $ at $t=0$ and then keep iterating in this way. There are $D \times K$ terms $ \psi(\Omega^{d},k) $ to compute at each update, costing $O(1)$ each. We have to repeat this $N$ times (once after the update of each $u_i=(u_{i1},\dots,u_{iK})$) for a total complexity of $N \times D \times K$. We have a similar complexity for updating the terms $\psi(\bar{\Omega}^{d}_{ik},k)$. As for the initialization of $ \psi^{(0)}(\Omega^{d},k) $, at $t=0$ we assume a fixed $u_{ik}=u_{k}$ for each node and calculate \cref{eqn:psiomega} analytically. Note that one can then randomly initialize the $u_{ik}$ and continue iterating using the updates above. We get
\be
 \psi^{(0)}(\Omega^{d},k) = \sum_{e \in \Omega^{d}}\prod_{j \in e}u_{k} = \sum_{e \in \Omega^{d}}u_{k}^{d} = \binom{N_{k}}{d} \, u_{k}^{d}\quad,
\ee
where $N_{k}$ is the number of nodes that have $u_{ik}>0$, i.e., the number of nodes initially in community $k$. 
With this formulation, the updates of the membership matrix become
\be\label{eqn:u3}
u_{ik} =  \f{\sum_{e\in \mathcal{E}}B_{ie} \,\rho_{ek} }{\sum_{d=2}^{D} w_{dk} \psi(\bar{\Omega}^{d-1}_{ik},k) } \quad.
\ee

We now compute the derivative of $\mathcal{L}(\rho, \theta)$ in $w_{dk}$:
\be
\f{\partial  \mathcal{L}}{\partial w_{dk} }= -\sum_{e \in \Omega^{d}} \, \prod_{j \in e} u_{jk} + \sum_{e \in \mathcal{E} | d_{e}=d}A_{e} \f{\rho_{ek} }{w_{dk}} \quad.
\ee
Setting this to zero, we get the updates:
\be\label{eqn:w}
w_{dk} = \f{\sum_{e \in \mathcal{E} | d_{e}=d}A_{e} \,\rho_{ek} }{\sum_{e \in \Omega^{d}}\, \prod_{j \in e} u_{jk}} =  \f{\sum_{e \in \mathcal{E} | d_{e}=d}A_{e} \,\rho_{ek}}{\psi(\Omega^{d},k)}\quad ,
\ee
which are computationally efficient and can be updated in parallel.

We describe the whole inference routine in Algorithm \ref{alg:EM}.

{\centering
	\begin{minipage}{.7\linewidth}
\setlength{\textfloatsep}{5pt}
\begin{algorithm}[H]
	\SetKwInOut{Input}{Input}
	\setstretch{0.7}
	\Input{hypergraph $A=\{A_{e}\}_{e \in \mathcal{E}}$, number of communities $K$.}
  	\BlankLine
	\KwOut{membership matrix $u=\rup{u_{ik}}$; affinity matrix $w=\rup{w_{dk}}$.}
	\BlankLine
	 Initialize $w$ and $u_{ik}=u_k$ at random.\\
	 Compute 
	 \be
     \psi^{(0)}(\Omega^{d},k) = \binom{N_{k}}{d} \, u_{k}^{d}\quad,
    \ee
	 
	 \BlankLine
	 Repeat until convergence:
	 \BlankLine
	\quad 1. Calculate $\rho$ (E-step) for $k,e \in \mathcal{E}$: 
    \be
    \rho^{(t)}_{ek} = \f{w^{(t-1)}_{d_{e}k} \prod_{i \in e} u^{(t-1)}_{ik}}{\sum_{k}w^{(t-1)}_{d_{e}k} \prod_{i \in e} u^{(t-1)}_{ik}} \nonumber
    \ee
    \BlankLine
	\quad 2. Update parameters $ \theta$ (M-step):  
	\BlankLine
	\quad \quad  
		i) for each pair $(d,k)$ update affinity matrix:
        \be
        w^{(t)}_{dk}  =  \f{\sum_{e \in \mathcal{E} | d_{e}=d}A_{e} \,\rho^{(t)}_{ek}}{\psi^{(t)}(\Omega^{d},k)}
        \ee
        \BlankLine
    	\quad \quad
		ii) for each pair ($i, k$):\\~\\
		\BlankLine
		\quad \quad \quad ii.i) for each pair ($d, k$) update
		\be
        \psi^{(t)}(\bar{\Omega}^{d}_{ik},k)  = \psi^{(t)}(\Omega^{d},k)  - u^{(t-1)}_{ik}\, \psi^{(t)}(\bar{\Omega}^{d-1}_{ik},k)
		\ee
		\BlankLine
		\quad \quad \quad ii.ii) update membership
		\be
        u^{(t)}_{ik} =  \f{\sum_{e \in \mathcal{E}}B_{ie} \,\rho^{(t)}_{ek} }{\sum_{d=2}^{D} w^{(t)}_{dk}\, \psi^{(t)}(\bar{\Omega}^{d-1}_{ik},k)} 
        \ee
        \BlankLine
        \quad \quad \quad ii.iii) for each pair ($d, k$) update
        \be
        \psi^{(t+1)}(\Omega^{d},k) = \bup{u_{ik}^{(t)}-u_{ik}^{(t-1)}} \psi^{(t)}(\bar{\Omega}^{d-1}_{ik},k) +  \psi^{(t)}(\Omega^{d},k) 
        \ee
	\quad \quad \quad
	\caption{\hymt:  EM algorithm}
		\label{alg:EM}
\end{algorithm}
  	\end{minipage}
\par
}

\subsection{Priors, regularization, and constraints}
So far, we infer the values of $\theta$ by following a maximum likelihood approach, which is equivalent to assuming uniform priors on the parameters. However, we can also posit non-uniform priors on the parameters and compute maximum a posteriori estimations. For instance, we may be interested in enforcing sparsity. To this aim, we can consider exponential distribution priors with parameters $\gamma_{u}$ and $\gamma_{w}$ for the parameters $u$ and $w$, respectively. This results in two added terms in the log-likelihood, giving:
\be
\mathcal{L}^{'} = \mathcal{L}  - \gamma_{u} \sum_{i,k} u_{ik} -\gamma_{w}\sum_{d,k}w_{dk}\quad.
\ee
Note, this is equivalent to a $L_{1}$-regularization on the values of $u$ and $w$. Following the same computations as before, we get the new updates differ only by a constant term added in the denominators, e.g., 
\be\label{eqn:wL1}
w_{dk} = \f{\sum_{e \in \mathcal{E} | d_{e}=d}A_{e} \,\rho_{ek} }{\gamma_{w}+ \sum_{e \in \Omega^{d}}\, \prod_{j \in e} u_{jk}} \quad.
\ee

Similarly, we can arbitrarily add constraints on the parameters. For instance, we can impose the membership vectors to be probability vectors, i.e., $\sum_k u_{ik}=1 \, \forall{i}$. Also in this case, it leads to a constant term added in the denominator of the updates of $u_{ik}$. In our numerical experiments, we run the model with and without constraints, and present the results of the model that performs the best.

\section{Hyperedge prediction and cross-validation}\label{app:auc}
We assess the performance of our model by measuring the goodness in predicting missing hyperedges. In these experiments, we use a 5-fold cross-validation routine: we divide the dataset into five equal-size groups (folds), selected uniformly at random, and give the models access to four groups (training data) to learn the parameters; this contains 80\% of the hyperedges. One then predicts the hyperedges in the held-out group (test set). By varying which group we use as the test set, we get five trials per realization. When we use the baseline \pwmt, the training and the test sets are the subsets extracted from the initial ones, containing only the hyperedges with $d_e = 2$. Instead, when we use the baseline \gmt, we train the model on the graph obtained from clique expansions of the hyperedges in the training set.

As a performance metric, we measure the area under the receiver-operator characteristic curve (AUC) on the test data, and the final results are averages over the five folds. The AUC is the probability that a random true positive is ranked above a random true negative; thus the AUC is 1 for perfect prediction, and 0.5 for chance. Since the set of all possible hyperedges is large, it is not possible to compute the AUC on the whole training and test sets; hence we proceed with samples. In detail, we fix the number of comparisons we want to evaluate, here $10^3$. We then sample $10^3$ values from the non-zero entries (where exist a hyperedge) of the sets, and we save the inferred hyperedge probabilities in a vector $R_1$. We sample the same number of values from the zero entries (where do not exist a hyperedge), keeping this set balanced with $R_1$ in terms of hyperedge size distribution. We save the inferred hyperedge probabilities of this set of entries in a vector $R_0$. We then make element-wise comparisons and compute the AUC as
$$\text{AUC}=\frac{\sum (R_1 > R_0) + 0.5 \sum (R_1 == R_0)}{|R_1|} \quad , $$
where $\sum (R_1 > R_0)$ stands for the number of times $R_1$ has a higher value than $R_0$ in the element-wise comparisons; and $|R_1|=|R_0|$ is the length of the vector, which is equal to the number of comparisons we fix.

To predict the existence of a hyperedge, we use different approaches according to the structure under analysis. For \hymt, the probability of a hyperedge is given by \cref{eqn:eq1}. For \gmt, instead, we compute the probability of a hyperedge as the product of the probabilities of each edge of its clique expansion to exist. That is, $P(A_e) = \prod_{(ij) \in e_2} P(A_{ij} > 0)$, where $e_2$ is the 2-combination set of the hyperedge $e$. Notice, all the single pairwise interactions have to exist, to have a probability of the hyperedge greater than zero. When evaluating \pwmt, we measure the AUC only on the subset of the test set containing edges, i.e., hyperedges with $d_e=2$. To perform a balanced
comparison in this case, we also measure the AUC for both \hymt\text{} and \gmt\text{} on this set (pairs), while still training on the whole train set. This provides information on the utility of large hyperedges to predict pairwise interactions.

\begin{table*}[!t]
    \centering
    \setlength{\tabcolsep}{6.5pt}
    \resizebox{0.6\textwidth}{!}{%
	\renewcommand{\arraystretch}{1.2}
    \begin{tabular}{{l}*{8}{r}}
    \toprule
  $c_{in}/c_{out}$  & $N$ & $E$ & $\avg k$ & s($k$) &  $\avg d$ & s($d$) & $D$ & $K$ \\
        \midrule
        $1$ & $500$ & $2230.7$ & $44.6$ & $24.2$ & $10.0$ & $5.4$ & $29.7$ & $3$\\ 
        $2$ & $500$ & $2097.6$ & $29.6$ & $16.9$ & $7.1$ & $3.8$  & $22.5$ & $3$\\ 
        $3$ & $500$ & $2013.4$ & $24.4$ & $13.7$ & $6.1$ & $3.1$  & $19.1$ & $3$\\ 
        $4$ & $500$ & $1963.3$ & $21.9$ & $12.5$ & $5.6$ & $2.9$  & $18.0$ & $3$\\ 
        $5$ & $500$ & $1911.9$ & $20.3$ & $11.6$ & $5.3$ & $2.7$  & $17.2$ & $3$\\ 
        $6$ & $500$ & $1878.2$ & $19.2$ & $11.2$ & $5.1$ & $2.6$  & $16.3$ & $3$\\ 
        $7$ & $500$ & $1866.1$ & $18.7$ & $11.0$ & $5.0$ & $2.5$  & $16.2$ & $3$\\ 
        $8$ & $500$ & $1835.0$ & $18.0$ & $10.5$ & $4.9$ & $2.4$  & $15.7$ & $3$\\ 
        $9$ & $500$ & $1820.3$ & $17.5$ & $10.4$ & $4.8$ & $2.4$  & $14.8$ & $3$\\ 
        $10$ & $500$ & $1818.2$ & $17.4$ & $10.0$ & $4.8$ & $2.4$  & $16.0$ & $3$\\ 
	\bottomrule          
    \end{tabular}}
	\caption{\label{tab:synOmega} \textbf{Summary of synthetic data with different strength of community structure.} Shown are the strength of the community structure~($c_{in}/c_{out}$), the number of nodes~($N$), number of hyperedges~($E$), mean node degree~($\avg k$), SD of node  degree~(s($k$)), mean hyperedge size~($\avg d$), SD of hyperedge size~(s($d$)), maximum hyperedge size~($D$), and number of communities~($K$). The values are averages over ten independent samples.}
\end{table*}

\section{Experiments with synthetic data}\label{app:synthetic}
The empirical data studied in the manuscript do not have ground truth labels, and therefore it is difficult to test the ability of the methods in recovering communities in hypergraphs. To this aim, we study the behavior of the models in synthetic data with known communities. We also use these data to provide a more precise estimate of the computational complexity of the methods. We generate hypergraphs by using the \texttt{dcsbm\_hypergraph} function inside the package \texttt{xgi}~\footnote[1]{\url{https://xgi.readthedocs.io/en/latest/api/generators/xgi.generators.nonuniform.html#module-xgi.generators.nonuniform}}, which generates binary synthetic hypergraphs by following a bipartite formalism. Notice that this data generating process differs from that of our generative model.

\subsection{Community detection}
To test the ability of the methods in community detection in hypergraphs, we generate data with different values of assortative structure. In detail, we fix $N=500$ nodes, $K=3$ communities and approximately $E=2000$ hyperedges. See \Cref{tab:synOmega} for a complete summary of the descriptive statistics of the data. In addition, this model takes in input an $\Omega$ matrix that regulates the number of connections within ($c_{in}$) and between ($c_{out}$) communities, and we generate data by varying the strength of the community structure, measured by the ratio $c_{in} / c_{out}$. We fix $c_{in}=2500$ and vary the ratio $c_{in} / c_{out} \in [1, \dots, 10]$, and for each value we generate ten independent samples. For this experiment,  we use two additional methods for comparison: \chodrow\ and \SC. The first is the generative model proposed in \cite{chodrow2021generative}, which generalizes the Louvain graph community detection method and assumes a symmetric partition function called All-Or-Nothing (AON) according to which edges are expected to lie fully within clusters. The second, instead, is the spectral method of \cite{zhou2006learning} and performs community detection in hypergraphs by using the eigenpairs of the hypergraph Laplacian. More specifically, the communities are detected with the run of the k-means algorithm on the subspace spanned by these eigenvectors. \Cref{fig:synOmega_nmi} shows results in terms of Normalized Mutual Information for the synthetic hypergraphs with three known communities. When $c_{in}=c_{out}$ the hypergraphs follow a random distribution and no community structure is induced; thus the communities are hard to detect. Conversely, the community structure becomes stronger as the ratio $c_{in}/c_{out}$ increases, and it becomes easier to detect communities. \hymt\text{} presents a clear pattern as its performance improves as the structure of the hypergraphs strengthens, and it significantly outperforms all the other methods, with the exception of \SC, which however benefits from having an inference routine similar to the generative process of the synthetic data. Conversely, \gmt\text{} shows low and constant results regardless the strength of the community structure, which may get lost with the clique expansions. Instead, the curve of \chodrow\text{} is flat around zero, illustrating the difficulty of the model to retrieve communities in synthetic data even when a strong assortativity is predominant. One explanation for its poor performance could rely on the assumptions behind the AON affinity function, which 
may be too strong and not appropriate to model this type of data. In fact, even though we have assortativity, this does not imply that \textit{all} of the nodes in a given hyperedge are likely to be in the same community, but rather the majority of them. AON may be too sensitive to them being all in agreement. Perhaps other types of partition functions as described in \cite{chodrow2021generative}  would be more appropriate, but they are also computationally prohibitive to run, thus making it impractical for our tests. In addition, \chodrow\text{} benefits from having strong ground truth cluster signal in the low-size hyperedges (e.g., pairs or triangles) and may struggle in other cases. This could also be another possible explanation since the synthetic data have several hyperedges of higher size. 

\begin{figure}[!ht]
	\centering
	\includegraphics[width=0.8\linewidth]{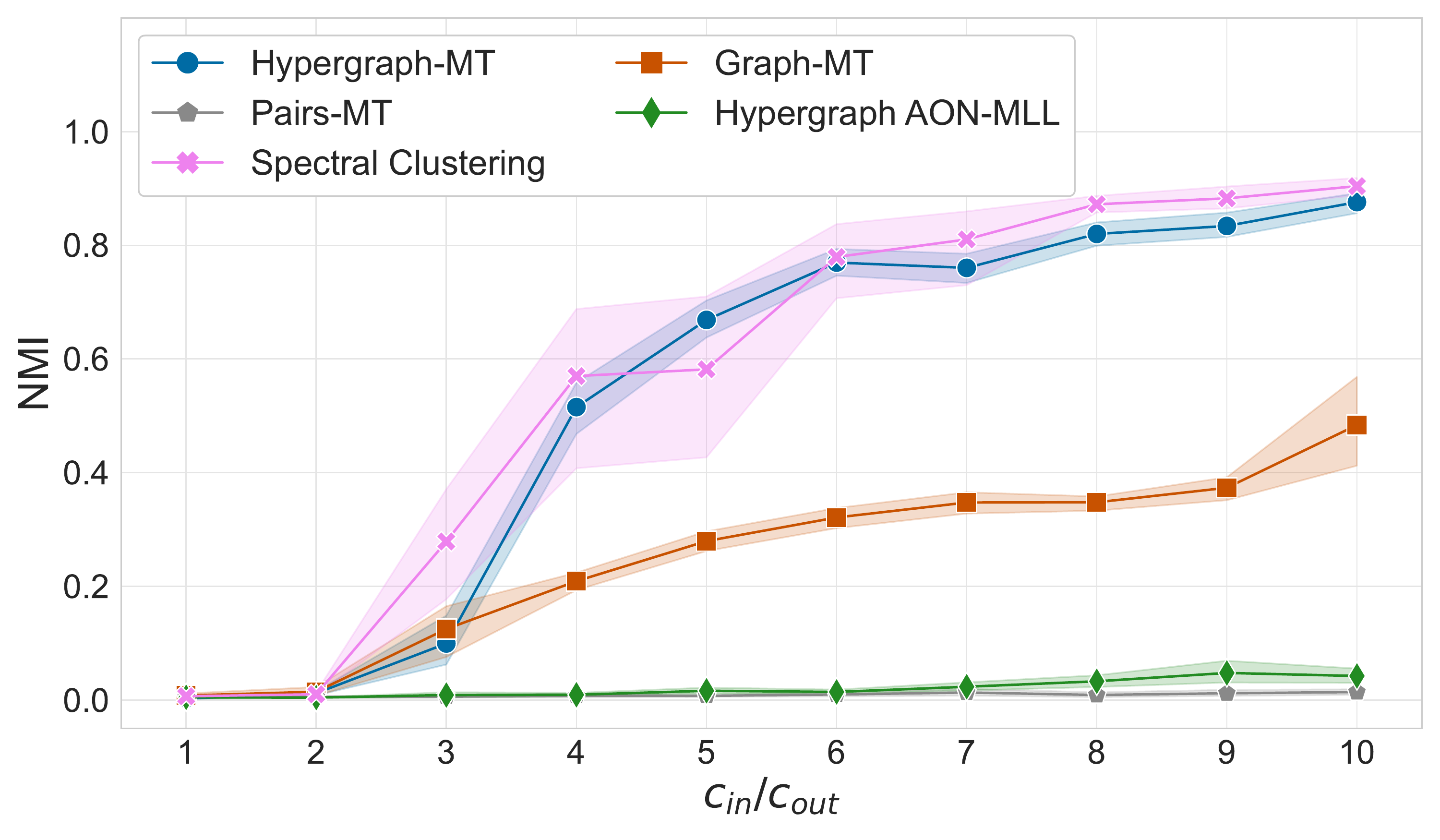}
	\caption{\label{fig:synOmega_nmi} \textbf{Community detection performance in synthetic hypergraphs with known communities.} We measure the Normalized Mutual Information in synthetic data generated by varying the strength of the community structure, measured by the ratio $c_{in}/c_{out}$. The descriptive statistics of the data are summarized in \Cref{tab:synOmega}, and the results are averages and standard deviations over ten independent samples. The plot shows how \hymt\ and \SC\  increase their performance as the community structure becomes stronger, while the other methods are not as robust. When a benchmark model with overlapping communities is used instead of one with hard communities, \hymt\ clearly outperforms \SC.}
\end{figure}

It is important to highlight that the synthetic data here used as ground truth were generated with an algorithm aimed at producing a planted partition from hard-membership communities~\footnote[1]{}. Indeed, both \chodrow\ and \SC\ are able to infer exclusively partitions where nodes are assigned to a single community. However, they lack the ability to capture the correct mesoscale organization of hypergraphs in more complex scenarios, where nodes can belong to more than one community at a time. To corroborate our statement, we have also investigated a different type of synthetic data, generated from a planted partition based on overlapping assignments of the nodes to multiple communities.
If we consider a very simple model  of overlapping community with two modules, where $25\%$ of the nodes belong exclusively to the first community ($u=[1, 0]$), $25\%$ of the nodes belong exclusively to the second community ($u=[0, 1]$), but $50\%$ of the nodes participate with equal strength to the two communities ($u=[0.5, 0.5]$), we find that \hymt\ clearly outperforms \SC. Indeed, comparing with ground truth, we obtain values of a cosine similarity $\text{CS}=0.97$ for \hymt\ and $\text{CS}=0.85$ for \SC. Moreover, the difference in performance between the two models becomes stronger when more complicated overlapping models are considered, even if we assign nodes predominantly to one community (a case which should favor hard-membership inference algorithms). For instance, if we consider a model where three groups of nodes of equal size are assigned community membership vectors of $u=[0.55, 0.35, 0.1$], $u=[0.05, 0.6, 0.35]$ and $u=[0.2, 0.2, 0.6]$ respectively, we obtain $\text{CS}=0.93$ for \hymt, but only $\text{CS}=0.67$ for \SC. A complete and detailed characterization of the hypergraph benchmark model used to generate ground truth data with overlapping communities will be provided in a separate manuscript currently in preparation.

\begin{table*}[!t]
    \centering
    \setlength{\tabcolsep}{6.5pt}
    \resizebox{0.55\textwidth}{!}{%
	\renewcommand{\arraystretch}{1.2}
    \begin{tabular}{*{8}{r}}
    \toprule
  $N$ & $E$ & $\avg k$ & s($k$) &  $\avg d$ & s($d$) & $D$ & $K$ \\
        \midrule
        $100$ & $141.6$ & $5.4$ & $3.0$ & $3.8$ & $1.7$ & $10.2$ & $3$\\ 
        $500$ & $690.7$ & $5.8$ & $4.0$ & $4.2$ & $2.0$  & $13.0$ & $3$\\ 
        $1000$ & $1355.3$ & $5.7$ & $3.9$ & $4.2$ & $2.0$  & $13.1$ & $3$\\ 
        $5000$ & $6768.6$ & $5.8$ & $4.1$ & $4.3$ & $2.1$  & $14.9$ & $3$\\ 
        $10000$ & $13521.9$ & $5.8$ & $4.1$ & $4.3$ & $2.1$  & $16.4$ & $3$\\ 
        \bottomrule          
    \end{tabular}}
	\caption{\label{tab:synN} \textbf{Summary of synthetic data with variable size.} Shown are the number of nodes~($N$), number of hyperedges~($E$), mean node degree~($\avg k$), SD of node  degree~(s($k$)), mean hyperedge size~($\avg d$), SD of hyperedge size~(s($d$)), maximum hyperedge size~($D$), and number of communities~($K$). The values are averages over ten independent samples.}
\end{table*}

\subsection{Time complexity}
In addition to the discussion in the manuscript, here we provide a proper assessment of the computational complexity of the various methods by running \hymt, \gmt, and \pwmt\text{} on synthetic data with variable size. We generate hypergraphs with average node degree $\avg k\approx 6$, average hyperedge size $\avg d \approx 4$, and an $\Omega$ matrix with entries $c_{in}=2 \, N$ and $c_{in}/c_{out}=10$. We vary the number of nodes $N\in[100, 500, 1000, 5000, 10000]$, and we generate ten independent samples for each different value. See \Cref{tab:synN} for a complete summary of the descriptive statistics of the data. \Cref{fig:timeN} displays the running times of the algorithms for one iteration, and shows a good scaling with hypergraph's size $N$, with \hymt\ being faster than \gmt\ across sizes. Notice that this computational complexity can be improved with a sparse implementation of the code, a task we leave for future work.

\begin{figure}[!htbp]
	\centering
	\includegraphics[width=0.8\linewidth]{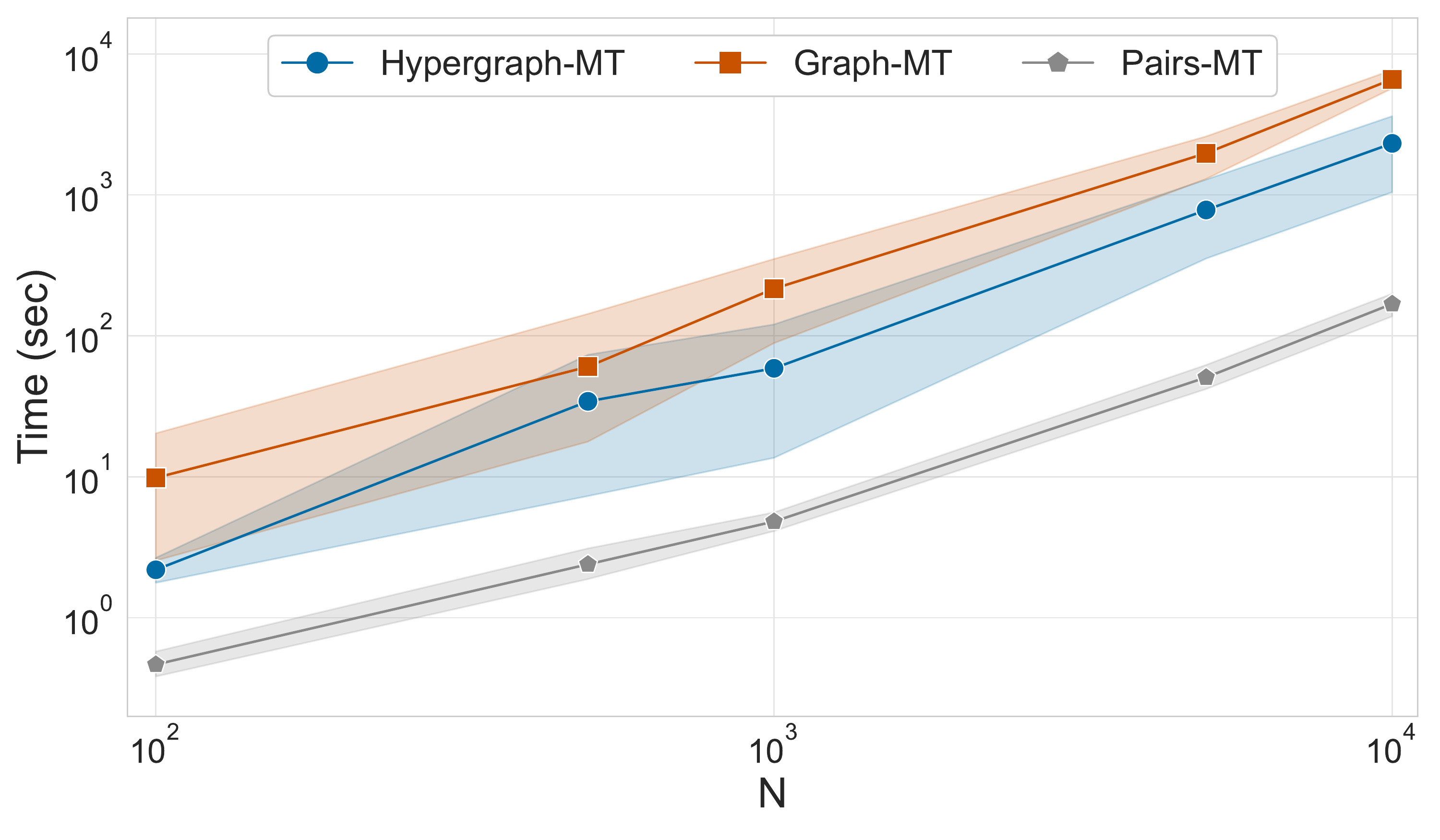}
	\caption{\label{fig:timeN} \textbf{Computational complexity in synthetic data with variable size.} We measure the running time for one realization of the methods in synthetic data generated by varying the number of nodes. The descriptive statistics of the data are summarized in \Cref{tab:synN}, and the results are averages and standard deviations over ten independent samples. The plot shows a good scaling with hypergraph's size $N$, with \hymt\ being faster than \gmt\ across sizes.}
\end{figure}

\section{Experiments with empirical data}\label{app:dataset}

In the manuscript, we analyze hypergraphs derived from empirical data from various domains, and we provide a summary of study datasets in Table I of the main text.  To perform the inference in these datasets, we need to choose the number of communities $K$. In general, $K$ can be selected using model selection criteria. For instance, one could evaluate the model's predictive performance--for example in the link prediction task--for varying numbers of communities, and then choose the best performing $K$. Here, for simplicity, we fix the number of communities $K$ equal to the number of classes of a node metadata, aiming to compare the resulting communities with this additional information. 

We first analyze four datasets collected by the SocioPatterns collaboration (\url{http://www.sociopatterns.org}), which describe human close-proximity contact interactions obtained from wearable sensor data. The High school dataset describes the interactions between students of nine different classrooms~\cite{mastrandrea2015contact}. In the Primary school, nodes are students and teachers and a hyperedge connects groups of people that were all jointly in proximity to one another~\cite{gemmetto2014mitigation, stehle2011high}. Also here, the number of communities reflects the classrooms to which each student belongs, and it includes an additional class for the teachers. The Workplace dataset contains the contacts of individuals of five different departments, measured in an office building in France~\cite{genois2015data}. Lastly, the Hospital hypergraph collects the interactions between patients, patients and health-care workers (HCWs) and among HCWs in a hospital ward in France~\cite{vanhems2013estimating}. The number of communities corresponds then to the number of roles in the ward.  

We then analyze the Gene-Disease dataset, that describes the gene-disease associations provided by expert curated resources (e.g., UNIPROT, CTI)~\cite{pinero2020disgenet}. Nodes correspond to genes, and each hyperedge is the set of genes associated with a disease. We keep only the genes with a non-nan value of the Disease Pleiotropy Index (DPI), a quantity that considers if the diseases associated with the gene are similar among them and belong to the same disease class or belong to different disease classes. We use this attribute to fix the number of communities because it may indicate the different behaviors of the genes in the datasets. Moreover, we keep hyperedges with size $2 \leq d_e \leq 25$.

The second case study in the manuscript presents the analysis of the Justice hypergraph constructed from the data in \url{http://scdb.wustl.edu/about.php}. This dataset records all the votes expressed by the justices of the Supreme Court in the U.S. from 1946 to 2019 case by case. Nodes correspond to justices, and each hyperedge is the set of justices that expressed the same vote in a case. The number of communities corresponds to the number of political parties, i.e., Democrat and Republican.

The following datasets have been downloaded from \url{https://www.cs.cornell.edu/~arb/data/}. We analyze hypergraphs created from U.S. congressional bill co-sponsorship data, where nodes correspond to congresspersons and hyperedges correspond to the sponsor and all cosponsors of a bill in either the House of Representatives (House bills) or the Senate (Senate bills)~\cite{fowler2006connecting, fowler2006legislative, chodrow2021generative}. We also use two datasets from the U.S. Congress in the form of committee memberships~\cite{stewart2008congressional, chodrow2021generative}. Each hyperedge is a committee in a meeting of Congress, and each node again corresponds to a member of the House (House committees) or a senator (Senate committees). A node is contained in a hyperedge if the corresponding legislator was a member of the committee during the specified meeting of Congress. In all these congressional datasets, the node labels give the political parties of the members, thus all of them have $K=2$. For these datasets, we run the model with different values of $D=2, \dots, 25$ and choose the best value among them. 

In addition to the congressional datasets, we analyze the Walmart hypergraph~\cite{amburg2020clustering}. Here, each node is a product, and a hyperedge connects a set of products that were co-purchased by a customer in a single shopping trip. We fix the number of communities equal to the product category labels. Lastly, we analyze the Trivago dataset~\cite{chodrow2021generative}. Nodes correspond to hotels listed at trivago.com, and each hyperedge corresponds to a set of hotels whose website was clicked on by a user of Trivago within a browsing session. For each hotel, the node label gives the country in which it is located, and we fix $K$ based on this information. For Walmart and Trivago, we consider a subset of the hypergraph to reduce the sparsity, as done in~\cite{chodrow2021generative}. The $c$-core of a hypergraph $\mathcal{H}$ is defined as the largest subhypergraph $\mathcal{H}_c$ such that all nodes in $\mathcal{H}_c$ have size at least $c$. For Walmart, we use the 3-core hypergraph, and for Trivago, we work with the 5-core hypergraph.

\clearpage
\section{Analysis of the Gene-Disease dataset} \label{app:geneprop}
\begin{figure}[!htbp]
	\centering
	\includegraphics[width=0.8\linewidth]{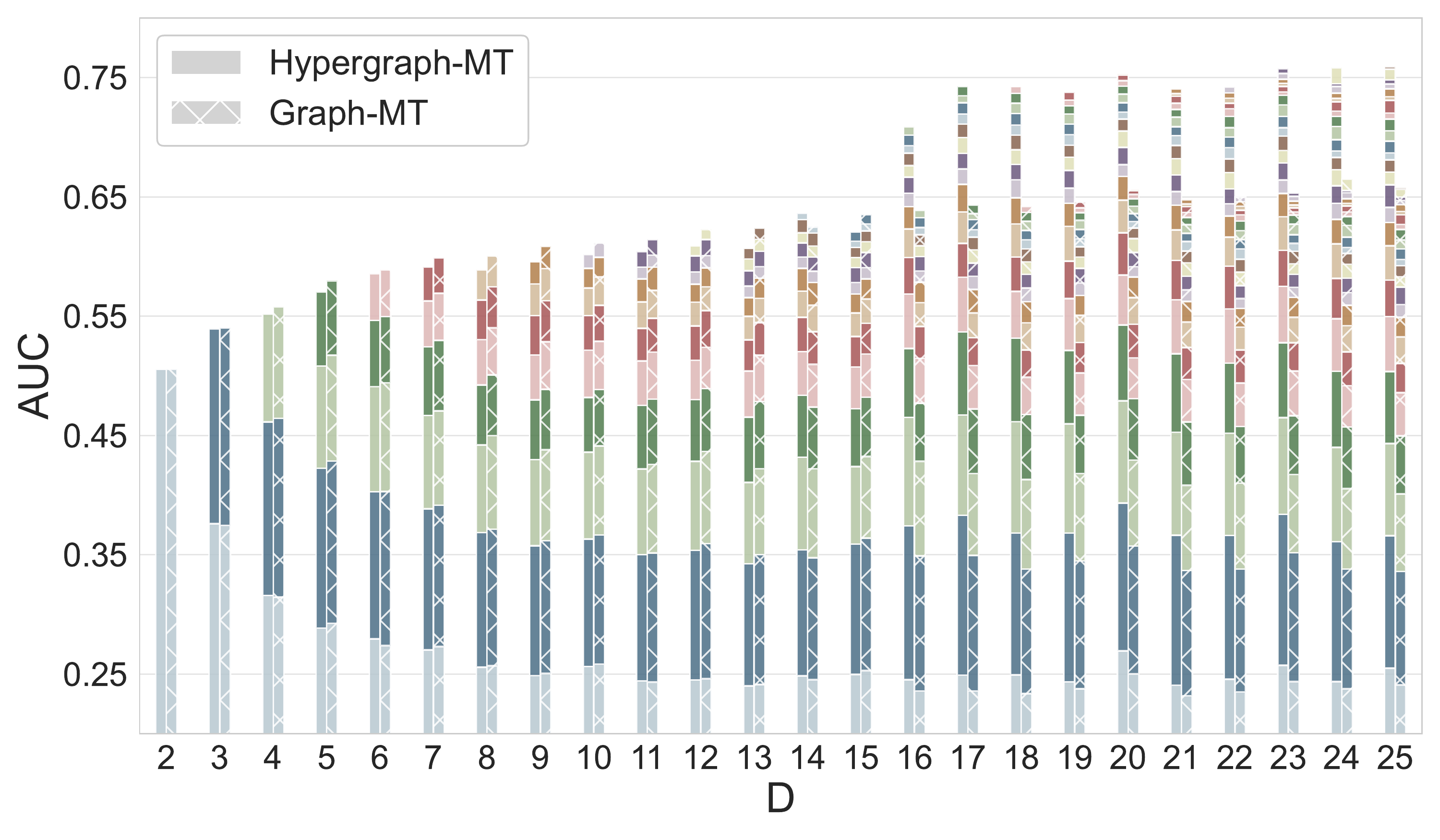}
	\caption{\label{fig:genedisease_auc_prop} \textbf{Cumulative hyperedge predictions in a Gene-Disease dataset.} We measure the AUC by varying the maximum hyperedge size $D$, and we plot the means over 5-fold cross-validation test sets. For each $D$, we show the cumulative performance for the different $2 \leq d \leq \dmax$. The plot shows how the model on the hypergraphs (Hypergraph-MT) outperforms the one using the graphs obtained by clique expansions (Graph-MT) beyond the shift around $\dmax = 15,16$. In particular, \hymt\text{} improves the predictive performance homogeneously across hyperedge sizes. Namely, it does not improve just in predicting the pairs-only, but also those of bigger sizes.}
\end{figure}

\end{widetext}

\end{document}